\documentclass[aip,reprint]{revtex4-1}

\usepackage{lipsum}
\usepackage{amsmath}
\usepackage{amssymb}
\usepackage{graphicx}
\usepackage{gensymb}
\usepackage{tabularx}
\usepackage{float}
\usepackage{xcolor}
\usepackage{mathtools}
\usepackage[capitalise]{cleveref}
\DeclareUnicodeCharacter{2009}{ }
\DeclareUnicodeCharacter{2192}{$\rightarrow$}
\DeclareUnicodeCharacter{2032}{'}
\DeclareUnicodeCharacter{03B1}{$\alpha$}
\DeclareUnicodeCharacter{03B3}{$\gamma$}
\restylefloat{table}
\graphicspath{ {Images/} }

\begin{document}
	\title{Measurements of dense fuel hydrodynamics in the NIF burning plasma experiments using backscattered neutron spectroscopy}
	\author{A. J. Crilly}\email{ac116@ic.ac.uk}
	\affiliation{Centre for Inertial Fusion Studies, The Blackett Laboratory, Imperial College, London SW7 2AZ, United Kingdom}
	\affiliation{I-X Centre for AI In Science, Imperial College London, White City Campus, 84 Wood Lane, London W12 0BZ, United Kingdom}
	\author{D. J. Schlossberg}
	\affiliation{Lawrence Livermore National Laboratory, Livermore, California 94550, USA}
	\author{B. D. Appelbe}
	\affiliation{Centre for Inertial Fusion Studies, The Blackett Laboratory, Imperial College, London SW7 2AZ, United Kingdom}
	\author{A. S. Moore}
	\affiliation{Lawrence Livermore National Laboratory, Livermore, California 94550, USA}
	\author{J. Jeet}
	\affiliation{Lawrence Livermore National Laboratory, Livermore, California 94550, USA}
	\author{S. M. Kerr}
	\affiliation{Lawrence Livermore National Laboratory, Livermore, California 94550, USA}
	\author{M. S. Rubery}
	\affiliation{Lawrence Livermore National Laboratory, Livermore, California 94550, USA}
	\author{B. Lahmann}
	\affiliation{Lawrence Livermore National Laboratory, Livermore, California 94550, USA}
	\author{S. O'Neill}
	\affiliation{Centre for Inertial Fusion Studies, The Blackett Laboratory, Imperial College, London SW7 2AZ, United Kingdom}
	\author{C. J. Forrest}
	\affiliation{Laboratory for Laser Energetics, University of Rochester, Rochester, New York 14623, USA}
	\author{O. M. Mannion}
	\affiliation{Sandia National Laboratories, Albuquerque, New Mexico 87185, USA}
	\author{J. P. Chittenden}
	\affiliation{Centre for Inertial Fusion Studies, The Blackett Laboratory, Imperial College, London SW7 2AZ, United Kingdom}
	
	\begin{abstract}
		The hydrodynamics of the dense confining fuel shell is of great importance in defining the behaviour of the burning plasma and burn propagation regimes of inertial confinement fusion experiments. However, it is difficult to probe due to its low emissivity in comparison to the central fusion core. In this work, we utilise the backscattered neutron spectroscopy technique to directly measure the hydrodynamic conditions of the dense fuel during fusion burn. Experimental data is fit to obtain dense fuel velocities and apparent ion temperatures. Trends of these inferred parameters with yield and velocity of the burning plasma are used to investigate their dependence on alpha heating and low mode drive asymmetry. It is shown that the dense fuel layer has an increased outward radial velocity as yield increases showing burn has continued into re-expansion, a key signature of hotspot ignition. Comparison with analytic and simulation models show that the observed dense fuel parameters are displaying signatures of burn propagation into the dense fuel layer, including a rapid increase in dense fuel apparent ion temperature with neutron yield.
	\end{abstract}

	\maketitle

	Recent inertial confinement fusion (ICF) experiments at the National Ignition Facility (NIF) have entered the `burning plasma'\cite{Kritcher2022,Zylstra2022} and `ignition'\cite{Abu2022,Zylstra2022ign,Kritcher2022ign} regimes. In these experiments, deuterium-tritium (DT) fuel is compressed to form a central fusing region, `hotspot', surrounded by a dense fuel layer, or shell. Within a burning plasma hotspot, alpha particle heating from DT fusion reactions dominates the input heating power from compression. Ignition marks the onset of a thermal instability started when alpha heating dominates all energy loss mechanisms. With ignition now achieved in the laboratory, understanding the hydrodynamic behaviour of the confining fuel layer and how the fusion burn propagates into the dense fuel shell is key in realising high energy gain ICF experiments. 
	
	The properties of the hotspot and dense fuel layer are coupled by hydrodynamics and energy exchange through thermal conduction, radiation transport and mass ablation. Within a burning hotspot, the temperature and fusion reaction rate continues to increase after maximal compression and into the re-expansion phase. The inertia of the dense fuel acts to confine the hotspot as it explodes radially and decompresses. The thermal gradient between hotspot and shell drives heat conduction which results in heating and mass ablation of the shell. 
	
	It follows that we expect the following characteristics of the dense fuel layer during neutron production in burning plasma and ignition regimes: lower fuel areal density, positive radial velocity (re-expansion) and elevated temperatures -- these signatures are unique to burning plasma experiments and will not be present in lower performing implosions. The areal density has been measured using the down-scattered-ratio (DSR) measurements\cite{Johnson2012} but the latter two are difficult to directly measure. Recent theoretical\cite{Crilly2020,Crilly2022} and experimental work\cite{Mannion2022} has shown that backscattered neutron spectroscopy can be used to measure the hydrodynamic conditions in the dense fuel layer during fusion burn. In this Letter we show, for the first time, inference of dense fuel velocity and apparent temperature in burning plasma ICF experiments which is critical to understanding the hydrodynamics and energy transport during burn propagation.

	When 14 MeV primary DT neutrons elastically scatter through 180$^o$ from ions they lose the largest fraction of their energy possible for a single scattering event. This produces an observable kinematic edge in the neutron spectrum. In ICF, ions have kinetic energies of order a keV which has a detectable effect on the scattering kinematics. Crilly \textit{et al.}\cite{Crilly2020} showed that the shape of the backscatter edges encodes information about the scattering ion velocity distribution. This information can be summarised as the scatter-averaged hydrodynamic quantities (fluid velocity, temperature, fluid velocity variance), analogous to the burn-averaging of the primary neutron spectra\cite{Munro2016,Appelbe2014}. Since the scattering rate is proportional to the product of the neutron flux and ion number density, scattering-averaged quantities are strongly weighted towards the dense fuel.
	
	Neutron time of flight (ntof) detectors at the OMEGA laser facility routinely measure the backscatter edge from tritium (or nT edge) to infer areal density\cite{Forrest2012}. A recent study\cite{Mannion2022} extended this work to successfully measure scatter-averaged hydrodynamic quantities from the nT edge. With the successful demonstration at OMEGA, the possibility of performing backscatter neutron spectroscopy at the NIF was investigated, specifically on the unique burning plasma experiments. The NIF ntof suite comprises 5 collimated lines of sight\cite{Moore2021}. Each line of sight includes a `Bibenzyl' scintillator detector which can measure a wide energy range, including the backscatter region. It was determined that the SPEC-SP detector (located at $\theta$ = 161.38$^o$, $\phi$ = 56.75$^o$) is gated sufficiently early to return a clean measurement of the nT edge. 
	
\begin{figure}[htp]
	\centering
	\includegraphics*[width=0.99\columnwidth]{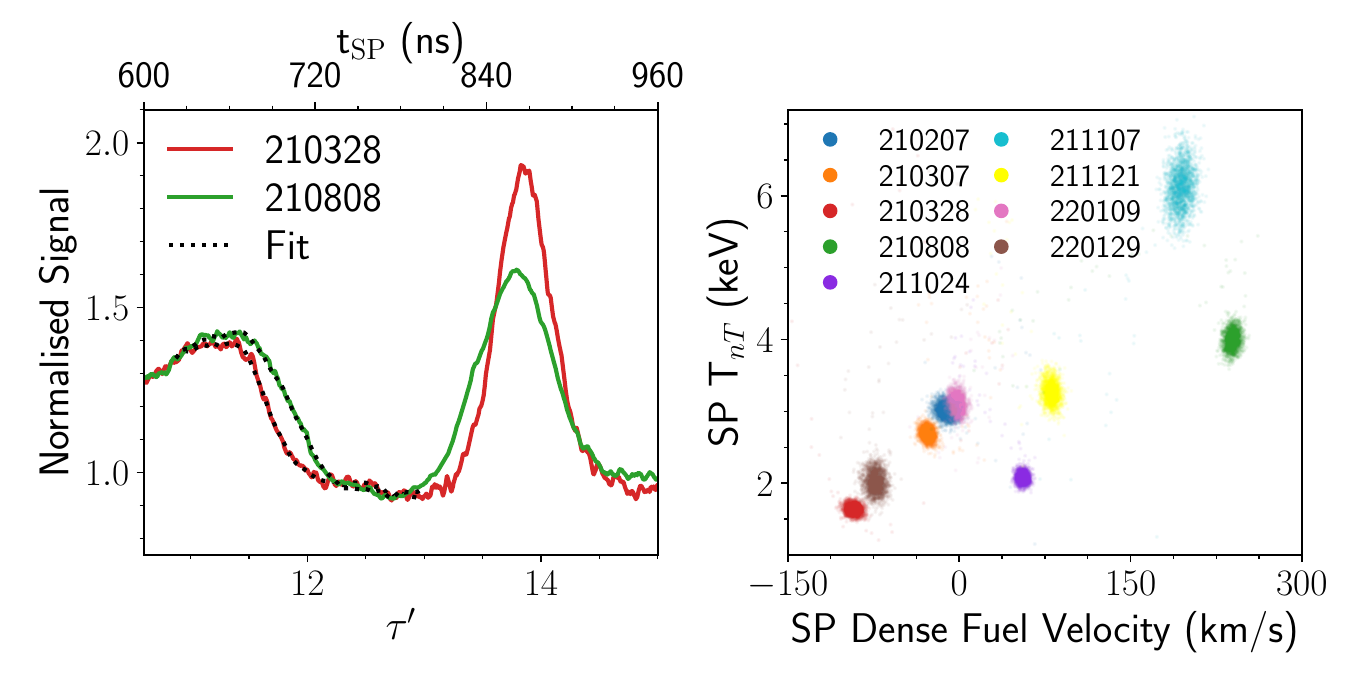}
	\caption{(Left) Neutron time of flight data from SPEC-SP normalised such that the signal amplitudes are equal above the nT edge to aid comparison between the two shots 210328 ($Y_n$ = 2 $\times$ 10$^{16}$, DSR = 3.4\%) and 210808 ($Y_n$ = 4 $\times$ 10$^{17}$, DSR = 3.2\%). Overplotted in the dotted black lines are the best fits using \cref{eqn:fitfunc}. (Right) The marginalised posterior values of dense fuel velocity and apparent ion temperature ($T_{nT}$) measured on the SP line of sight, as represented by point clouds drawn from a subset of the MCMC chain. For each experimental dataset, $\sim$1 million forward fit evaluations were used in the MCMC posterior distribution calculation.}
	\label{fig:ntof_data_selected}
	\end{figure}

	Backscattered neutron spectroscopy presents a novel challenge on the NIF. Similar to the OMEGA experiments, a forward fit methodology was used based on Mohamed \textit{et al.}\cite{Mohamed2020}. However, the higher areal densities increase the degree of multiple scattering and attenuation requiring a novel model to describe the edge spectral shape. A 6 parameter analytic model was devised for the energy spectrum based on a linear expansion of the cross section about the backscatter energy and a linear background (see supplementary material). Combining with the detector sensitivity, Jacobian and instrument response function allowed a forward fit of nToF data:
	\begin{align}
		I(\tau') &\propto \left[a(\tau) s(\tau) \frac{dN}{dE} \frac{dE}{d\tau}\right] \otimes R(\tau'-\tau,\tau) \ , \label{eqn:fitfunc}\\
		\tau &= \frac{t_n c}{d} = \frac{c}{v_n} \ ,
	\end{align}
	where $\tau$ is the neutron arrival time normalised by the photon arrival time, $\tau'$ is the normalised recorded signal time, $I$ is the measured detector signal, $a$ is the beamline attenuation model, $s$ is the detector sensitivity model and $R$ is the instrument response function. The detector models that were developed\cite{Hatarik2015} for the DD peak can be applied in the nT edge fits. The fit parameters of the energy spectrum, $dN/dE$, measure the scatter-averaged fluid velocity and apparent ion temperature. The measured hotspot and isotropic velocity, and apparent ion temperature projected along SPEC-SP were used to account for the effect of the shifted and broadened DT peak on the backscatter edge\cite{Crilly2020} and their uncertainties were included in edge parameter inference. A fit region between 2.8 and 4.0 MeV was constrained using synthetic neutron spectra from 1D radiation-hydrodynamics simulations for various levels of alpha heating\cite{Crilly2022}. The lower limit of the fit region is set by the exclusion of the DD peak and the upper limit extends beyond the full width of the edge to capture the edge jump height. The nToF data from two shots (210328 and 210808) are shown in \cref{fig:ntof_data_selected}. These shots have a large difference in yield, $\sim$ 2 $\times$ 10$^{16}$ (55 kJ) compared to $\sim$ 4 $\times$ 10$^{17}$ (1.35 MJ). Differences in the backscatter edge and DD peak spectra are clearly visible. The nT edge from shot 210808 appears at a later time suggesting an expanding dense fuel layer. It is also broader than 210328 suggesting a higher ion temperature or fluid velocity variance in the shell. Best fits to the data are found by minimising a $\chi^2$ loss function.
	
	A total of 9 burning plasma experiments were analysed, using the SPEC-SP nToF data for the backscatter analysis. A Markov Chain Monte Carlo (MCMC) algorithm\cite{emcee} was used to find the optimal fitting parameters and random uncertainties, the results of which are shown in \cref{fig:ntof_data_selected} -- systematic uncertainties are discussed in the supplementary material. In the following discussion we investigate correlations between the backscatter parameters, neutron yield and hotspot velocity along SPEC-SP and relate these to the physics of burning plasmas. 
	
	To aid with interpretation of the experimental results, a set of 30 1D radiation hydrodynamics Chimera simulations were used to investigate the trends due to ignition and burn propagation. Chimera is an Eulerian radiation magnetohydrodynamics code with an alpha heating model, details of ICF implosion modelling with Chimera can be found in the literature\cite{Chittenden2016,Walsh2017,McGlinchey2018,Tong2019,Crilly2022,SpK2022}. The simulation set was performed at a fixed hydrodynamic scale (1014$\mu$m inner radius) and HDC design but a uniform mix fraction of carbon in the fuel was varied to modify the radiative losses and consequently the yield.
	
\begin{figure}[htp]
	\centering
	\includegraphics*[width=0.99\columnwidth]{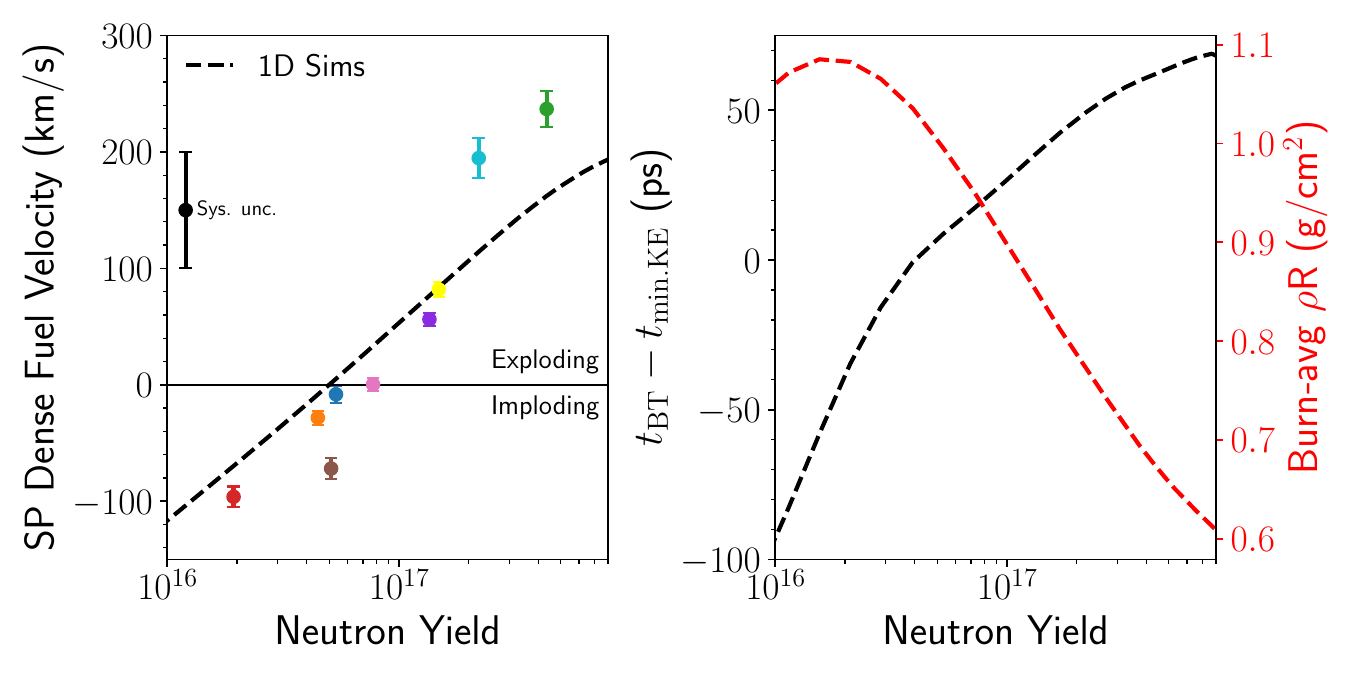}
	\caption{(Left) Dense fuel velocity against neutron yield with 1 standard deviation random uncertainty error bars. Also shown is the calculated systematic uncertainty, displayed as the black error bar, and 1D Chimera simulation predictions as a black dashed line. Note positive velocity means radial expansion of the dense fuel. (Right) From the 1D simulation dataset, the time difference between time of peak neutron production or `bang time', $t_{\mathrm{BT}}$, and time of minimum fuel kinetic energy, $t_{\mathrm{min. KE}}$, is shown as a function of neutron yield in black. The neutron yield at which $t_{\mathrm{BT}}$ equals $t_{\mathrm{min. KE}}$ also coincides with the dense fuel velocity approaching zero i.e. stagnation. The burn-averaged fuel areal density is shown in red and shows a peak when fusion burn occurs during stagnation. For higher yields the fuel is decompressing during burn and thus the burn-averaged areal density decreases.}
	\label{fig:densefuelvelocity}
	\end{figure}

	Firstly, it is found that there is a positive correlation between neutron yield and the dense fuel fluid velocity as shown in \cref{fig:densefuelvelocity}. As neutron yields increase above $\sim$ 5 $\times$ 10$^{16}$, we see the dense fuel start to explode (positive radial velocity). This reflects two changes introduced by significant alpha heating: a shift of peak neutron production (bang time) to later times and increased hotspot pressure. These can be understood in terms of the hotspot power balance. During implosion mechanical work and alpha heating are energy sources while thermal conduction and radiative losses are energy sinks\cite{Hurricane2021}. As alpha heating is increased, the total hotspot power can remain positive towards stagnation when the mechanical work vanishes. With sufficiently high alpha heating, the total hotspot power can remain positive during the explosion phase when mechanical work is an energy sink. The measured dense fuel fluid velocity will correlate with the mechanical work at bang time and thus reflects the changes in hotspot power balance due to alpha heating. Alpha heating both maintains positive hotspot power to later times and increases the hotspot pressure. This increased pressure drives more rapid expansion and decompression of the shell after stagnation.
	
\begin{figure*}[htp]
	\centering
	\includegraphics*[width=0.9\textwidth]{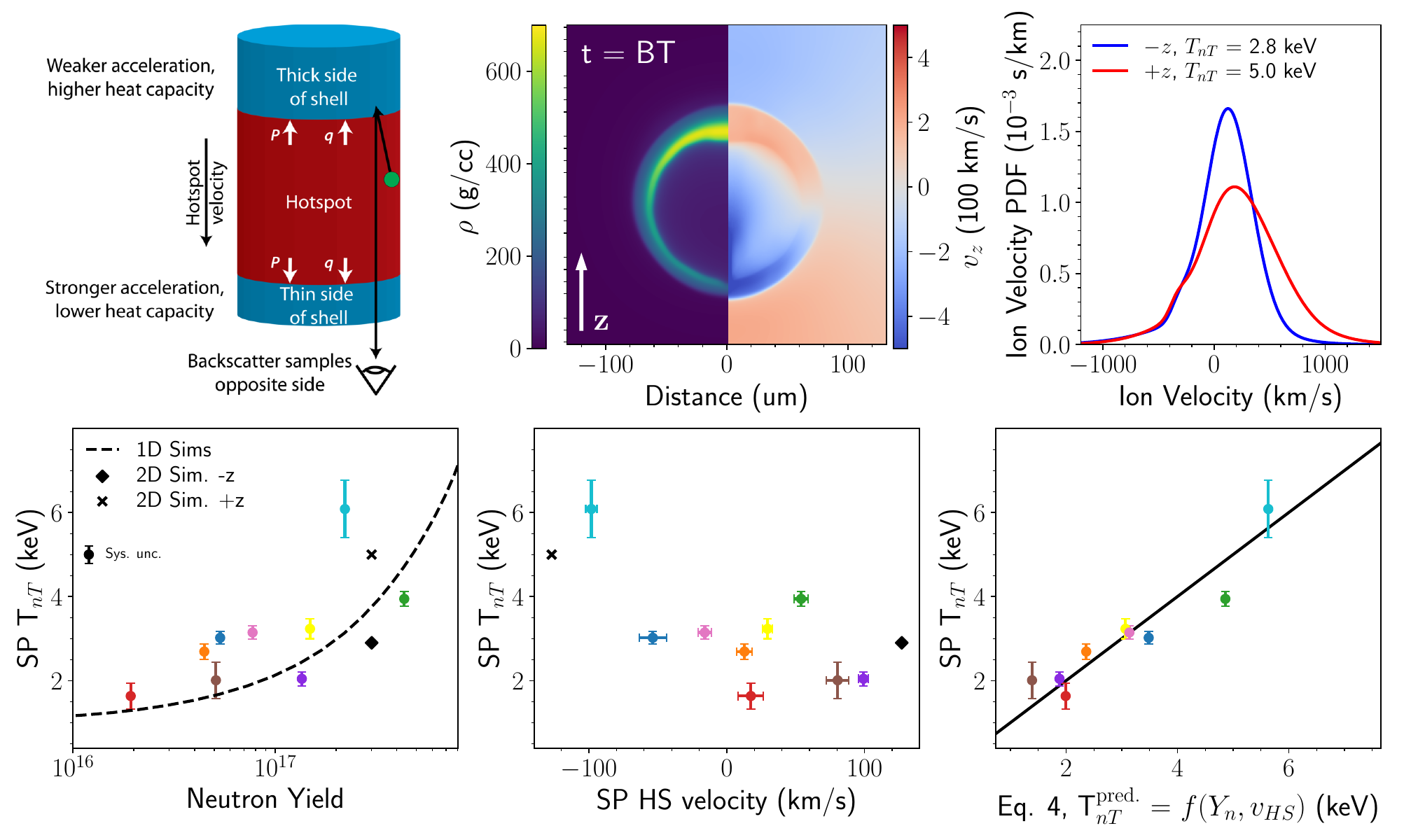}
	\caption{(Top Left) The asymmetric piston abstraction of a mode 1 asymmetry. The hotspot pressure, $P$, and heat flux, $q$, impart differential acceleration and heating in the dense fuel due to the unequal masses of the pistons, inducing anisotropy in the $T_{nT}$ measurement. (Top Centre) Mass density and the $z$ component of the fluid velocity from a Chimera simulation using N210808 parameters with a mode 1 drive asymmetry applied. (Top Right) The scattering ion velocity probability distribution function for the $+z$ and $-z$ lines of sight, the hotspot flow is away from and towards these detectors respectively. The $-z$ line of sight observed backscattered neutrons from the thick side of the shell. Since the thick side of the shell is more massive, it is colder, and expanding and accelerating slower than the thin side. This is reflected in the scattering ion velocity PDFs. Neutron transport calculations were used to evaluate $T_{nT}$ for each line of sight. (Bottom Left) $T_{nT}$ against neutron yield with 2D Chimera simulation results plotted as black symbols. (Bottom Centre) $T_{nT}$ against hotspot velocity projected along the SPEC-SP line of sight. (Bottom Right) Measured $T_{nT}$ compared to the predicted $T_{nT}$ from the empirical model given in \cref{eqn:TnTfit}.}
	\label{fig:Mode1Analysis}
	\end{figure*}

	Secondly, it is found that the dense fuel apparent ion temperature, $T_{nT}$, correlates with both neutron yield and the hotspot velocity projected along the SPEC-SP line of sight as shown in \cref{fig:Mode1Analysis}. We hypothesize that these correlations are due to two separate physical phenomena. The first of which is due to the effects of burn propagation. With increased yield, we expect increased hotspot heating to drive heat transport in the form of electrons and alpha particles into the dense fuel. The dense fuel is heated and ablated into the hotspot, reaching thermonuclear temperatures. As a larger fraction of the fuel mass is heated by burn propagation\cite{Tong2019} we expect a corresponding increase in the scatter-averaged thermal temperature. Alpha heating also increases the hotspot pressure, $P$, and thus produces large acceleration, $a$, of the dense fuel radially outwards given $a = P/\rho R_{\mathrm{shell}}$. Over the duration of burn, the large acceleration will cause large fluid velocity variance in the dense fuel. Since the apparent $T_{nT}$ is sensitive to both the thermal temperature and fluid velocity variance of the dense fuel, we expect $T_{nT}$ to increase with increased alpha heating and therefore fusion yield. As discussed in previous work\cite{Crilly2022}, this sensitivity makes the backscatter edge a unique diagnostic of burn propagation into the dense fuel. 

	The backscatter edge is also sensitive to anisotropy in the dense fuel hydrodynamic conditions. As we are observing neutrons which scattered through 180$^o$, the diagnostic is `imaging' the back side of the implosions with respect to the detector line of sight. Low mode asymmetries have been identified as a common yield degradation mechanism in current ICF experiments\cite{Hurricane2020,Rinderknecht2020}. In particular, mode $L$=1 asymmetries induce centre of mass motion of the imploding capsule which can be diagnosed by Doppler shifts in the primary neutron spectra\cite{Hatarik2018}. This can be reconstructed into a hotspot velocity vector, showing the magnitude and direction of the asymmetry. Radiation hydrodynamics simulations have shown that mode 1 drive asymmetries produce co-aligned areal density asymmetries\cite{Spears2014}.
	
	Following the abstraction of Hurricane \textit{et al.}\cite{Hurricane2020}, we can consider the mode 1 system as two unequal mass pistons compressing a central hotspot and inducing a centre of mass velocity. We can predict the effect on $T_{nT}$ by considering the qualitative behaviours of the thermal temperature and fluid velocity of the pistons. Firstly, the more massive side will have a higher heat capacity and therefore remain colder when the hotspot begins to transport heat into it. Secondly, the more massive piston will experience a lower acceleration and thus have a lower fluid velocity variance. The opposite is then true of the less massive side, producing anisotropy in the backscatter measurement. Combining these effects, one expects to measure a lower $T_{nT}$ on a detector where the hotspot is flowing towards it. We can produce a theoretical prediction of the anisotropy in $T_{nT}$ due to differential acceleration of the thick and thin sides of the shell (see supplementary material for derivation):
	\begin{align}
		\Delta T_{nT} & \approx  -0.8 \mbox{ keV} \frac{v_{imp}}{400 \mbox{ km/s}}\frac{v_{HS}}{100 \mbox{ km/s}} \ ,
	\end{align}
	where $v_{imp}$ is the implosion velocity and $v_{HS}$ is the hotspot velocity. 
	
	Investigating these arguments further, a 2D radiation hydrodynamics simulation of a capsule driven with a mode 1 drive asymmetry was performed with the code Chimera\cite{Chittenden2016}. The N210808 capsule parameters and tuned 1D drive were used with an additional constant mode 1 drive asymmetry included. A summary of the findings are given in \cref{fig:Mode1Analysis}. A simulated burn-averaged hotspot velocity magnitude of 127 km/s was achieved, which is similar to the maximum hotspot velocity observed in the subset of experiments considered in this work. Simplified neutron transport calculations\cite{Crilly2018} calculated scattering ion velocity distributions from which we obtain values for the scatter-averaged quantities parallel and anti-parallel to the hotspot velocity i.e. of the thin and thick sides of the shell. It was found that the $T_{nT}$ was largest for the thin side of the shell, 5.0 keV compared to 2.8 keV for the thick side of the shell - following the intuition of the asymmetric piston model. Additionally, the thin side of the shell was found to be exploding at 220 km/s compared to 90 km/s of the thick side.
	
	From theory, simulation and experiment, it is clear that the measured $T_{nT}$ depends on both the level of burn propagation and magnitude of mode 1 asymmetry. If we assume these effects are decoupled, we can construct an empirical model of the measured $T_{nT}$ using the 1D simulation results, the asymmetric piston model of mode 1 anisotropy and an additional degree of freedom to capture higher order effects. We can then fit the following model to the experimental data:
	\begin{equation}\label{eqn:TnTfit}
		T_{nT}^{\mathrm{meas.}} = T_{nT}^{1D}(Y_n^{\mathrm{meas.}})+\alpha_{\mathrm{LM}}v_{HS}^{\mathrm{meas.}}+\alpha_{\mathrm{HM+sys.}} \ ,
	\end{equation}
	where the best fit parameters are $\alpha_{\mathrm{LM}} = -1.5 \pm 0.3$ keV/(100 km/s), which can be interpreted as the degree of shell anisotropy induced by low mode asymmetries, and $\alpha_{\mathrm{HM+sys.}} = 0.8 \pm 0.2$ keV, which can be interpreted as the effect of unresolved high mode asymmetries and/or systematic uncertainty in the measurements. Previous backscatter spectroscopy results on OMEGA also identified an unresolved high mode component to the apparent dense fuel temperature attributed to hydrodynamic instabilities in the imploding shell\cite{Mannion2022}. For the low mode component, the 2D Chimera simulation gives a $\alpha_{\mathrm{LM}} = -0.8$ keV/(100 km/s) which agrees with the asymmetric piston model prediction. Experimentally, we infer an increased dense fuel anisotropy due to low mode asymmetries compared to these predictions. A potential physical explanation is that mode 1 anisotropic acceleration produces anisotropic growth of hydrodynamic instabilities\cite{Clark2019}. Therefore, there will be a directional dependence to the amplitude of high mode asymmetries which are not resolved in the Chimera simulation or asymmetric piston model. 
	
	In conclusion, we present novel neutron backscatter spectroscopy measurements of the dense fuel hydrodynamic conditions at the NIF. These measurements give unique insight into the hydrodynamics of the dense fuel in burning plasma implosions. As neutron yield increases, it is seen that the dense fuel is burning during re-expansion, is exploding faster and has a raised apparent ion temperature. These measured trends are consistent with simulation predictions of the hotspot ignition and propagating burn regimes. Improving agreement between simulation and experiment can be used to constrain models of burn propagation and identify the dominant hydrodynamic and energy transport phenomena. Daughton \textit{et al.} showed the enthalpy flux between shell and hotspot is sensitive to both the dense fuel temperature and the relative importance of electron thermal conduction, alpha particle heating and transport, and potentially fusion neutron heating\cite{Daughton2022}. The anisotropy in dense fuel conditions created by mode 1 asymmetries is also seen to affect the backscatter spectra. This was identified through a trend between hotspot velocity and apparent dense fuel temperature. A consistent trend was found with the asymmetric piston model\cite{Hurricane2020} and in a 2D radiation hydrodynamic simulation with a mode 1 drive asymmetry. Therefore, a 3D picture of the dense fuel conditions can be achieved when more ntof lines of sight are used to measure the backscatter spectra. Understanding burn propagation into the dense fuel is key in achieving high gain in inertial confinement fusion and the results of this paper show that the backscatter edges encode important information on shell conditions and burn propagation.

	\section*{Acknowledgements}
	Sandia National Laboratories is a multimission laboratory managed and operated by National Technology \& Engineering Solutions of Sandia, LLC, a wholly owned subsidiary of Honeywell International Inc., for the U.S. Department of Energy’s National Nuclear Security Administration under contract DE-NA0003525. 
	
	This work was performed under the auspices of the U.S. Department of Energy by Lawrence Livermore National Laboratory under Contract No. DE-AC52-07NA27344. This article (No. LLNL-JRNL-846265-DRAFT) was prepared as an account of work sponsored by an agency of the U.S. government. Neither the U.S. government nor Lawrence Livermore National Security, LLC, nor any of their employees make any warranty, expressed or implied, or assume any legal liability or responsibility for the accuracy, completeness, or usefulness of any information, apparatus, product, or process disclosed, or represent that its use would not infringe privately owned rights. Reference herein to any specific commercial product, process, or service by trade name, trademark, manufacturer, or otherwise does not necessarily constitute or imply its endorsement, recommendation, or favoring by the U.S. government or Lawrence Livermore National Security, LLC. The views and opinions of authors expressed herein do not necessarily state or reflect those of the U.S. government or Lawrence Livermore National Security, LLC,and shall not be used for advertising or product endorsement purposes.
	
	This project is supported by the Eric and Wendy Schmidt AI in Science Postdoctoral Fellowship, a Schmidt Futures program.

	\section*{Supplementary Material}
	\subsection{Spectral model}
	At NIF scale areal densities both multiple scattering and differential attenuation are non-negligible. Therefore, devising an \textit{ab-initio} model of the backscatter spectrum is challenging. We must make a number of simplifying assumptions in order to construct a suitable fitting model for the nT backscatter edge:
	\begin{itemize}
		\item The attenuated nT single scatter spectrum can be expanded to linear order in the vicinity of the edge.
		\item The distribution of backscatter energies, due to both primary neutron and scattering ion velocity variations, is Gaussian. This has been justified in previous studies\cite{Crilly2020,Mannion2022}.
		\item The background from other scattering processes, including multiple scattering, can be sufficiently described by a linear function of energy. 
	\end{itemize}
	Constructing the resulting neutron spectrum using these approximations yields:
	\begin{subequations}
		\begin{align}
		\frac{dN}{dE} &= \mathrm{nT} + \mathrm{Background} \\
		&= \int (c_1'+c_2' (K_n-K_n^*)) \\
		&\Theta(K_n-K_n^*) \exp\left[-\frac{(K_n^*-\mu_ E)^2}{2\sigma_E^2}\right] dK_n^* \nonumber \\
		&+ c_3' K_n + c_4' \ ,
		\end{align}
	\end{subequations}
	where $K_n$ is the scattered neutron kinetic energy, $K_n^*$ is the nT backscatter energy and the Heaviside function, $\Theta$, enforces the kinematic endpoint of the nT scattering. Performing the integration, we arrive at the final model:
	\begin{subequations}
		\begin{align}
		\frac{dN}{dE} &= c_1 \mathrm{erf}\left(\frac{K_n-\mu_E}{\sqrt{2}\sigma_E}\right) \\
		&+ c_2 \left[\sqrt{\frac{2}{\pi}}\sigma_E\exp\left(-\frac{(K_n-\mu_E)^2}{2\sigma_E^2}\right)\right. \\ &\left.+(\mu_E-K_n)\mathrm{erf}\left(\frac{K_n-\mu_E}{\sqrt{2}\sigma_E}\right)\right] \nonumber\\
		&+ c_3 K_n + c_4 
		\end{align}
	\end{subequations}
	where $\mu_E$ and $\sigma_E$ are the edge shape parameters and $c_i$ are amplitude coefficients, giving a total of 6 free parameters.
	
	The fitting parameters $\mu_E$ and $\sigma_E$ can be related to scatter-average quantities\cite{Crilly2020} by manipulation of the backscatter kinematics equation (assuming non-relativistic ion velocity):
	\begin{equation}\label{eqn:Kn_BS}
	K_n \approx \frac{(A_i-1)^2}{A_i^2+2A_i\gamma_n' +1}K_n' + 2 \frac{A_i(A_i^2-1)(A_i+\gamma_n')}{(A_i^2+2A_i\gamma_n' +1)^2}p_n' v_i'
	\end{equation}
	where $K$, $\gamma$, $p$ and $v$ are the kinetic energy, Lorentz factor, momentum and velocity with species denoted by subscript, primed and unprimed denoting pre- and post-collision values respectively and $A_i$ is the ratio of the ion to neutron mass. From the above, we find\cite{Crilly2022}:
	\begin{subequations}
		\begin{align}
		\langle v_i' \rangle &\approx  \frac{(A_i^2+2A_i\gamma_n' +1)^2}{2A_i(A_i^2-1)(A_i+\gamma_n')}\frac{\mu_E - \langle K_{nT,0} \rangle}{\langle p_n'\rangle} \\
		T_{nT} &\approx \frac{(A_i^2+2A_i\gamma_n' +1)^3}{8A_i(A_i+1)^2(A_i+\gamma_n')^2} \frac{\sigma_{E,nT}^2}{\langle K_{nT,0} \rangle} \\
		\langle K_{nT,0} \rangle &=  \frac{(A_i-1)^2}{A_i^2+2A_i\gamma_n' +1}\langle K_n' \rangle \ , \\
		\sigma_{E,nT}^2 &= \sigma_E^2- \left(\frac{(A_i-1)^2}{A_i^2+2 A_i \gamma_n'+1}\right)^2 \mathrm{Var}(K_n')
		\end{align}
	\end{subequations}
	In these calculations we neglect the variance in the pre-collision Lorentz factor as this is order $\mathrm{Var}(K_n')/(M_n^2 c^4)$. The accuracy of these approximate analytic expressions was tested numerically using Monte Carlo sampling of test distributions and shown to have errors of $\sim$ 5 km/s and $\sim$ 50 eV for $\langle v_i' \rangle$ and $T_{nT}$ respectively. It is important to note that the pre-collisions quantities include any directional changes to the DT primary spectral moments ($\langle K_n' \rangle$ and $\mathrm{Var}(K_n')$), e.g. from hotspot velocity flows. 
	\subsection{Systematic uncertainty quantification}
	Relevant systematic uncertainties can be separated in model and measurement effects. Absolute timing uncertainty will affect the velocity inference more than temperature, which is a differential measurement. The uncertainty in dense fuel velocity is $\approx 22 \ \mbox{km/s} \ (\Delta t /\mbox{1 ns}) ( \mbox{20 m} / d)$ for a given timing uncertainty, $\Delta t$, and distance to detector, $d$. Hatarik \textit{et al.}\cite{Hatarik2018} report a total timing uncertainty of $\sim$ 0.1 ns on the NIF nToF suite. The model uncertainties include uncertainty in the IRF, sensitivity and spectral models. The peak and FWHM of the IRF at 3.5 MeV are $\sim$ 3.5 ns and $\sim$ 6 ns respectively. An assumed 10\% uncertainty in these IRF parameters introduce errors of $\sim$ 8 km/s and $\sim$ 20 eV to the edge parameters. The spectral model assumes a Gaussian form for the scattering ion velocity distribution which neglects profile effects\cite{Crilly2020} and invokes an empirical form for the background neutron signal. Fits to synthetic neutron spectra with known dense fuel conditions\cite{Crilly2022} show these approximations have dense fuel properties systematic errors of $\sim$ 45 km/s and $\sim$ 200 eV at NIF scale areal densities. Additionally, approximations made in the scattering kinematics introduce model errors of $\sim$ 5 km/s and $\sim$ 50 eV for the dense fuel velocity and temperature respectively, as discussed in the section above. Combining all contributions in quadrature, the total systematic uncertainties were found to be 50 km/s and 210 eV for the dense fuel velocity and apparent temperature respectively.
	\subsection{Asymmetric piston model prediction of $T_{nT}$ anisotropy}
	Using the analytic results of Hurricane \textit{et al.}\cite{Hurricane2020}, one can derive the change of velocity over the hotspot confinement time for a piston with areal density $\rho R$:
	\begin{equation}
	a_{stag} \Delta t = \frac{P_{stag}\Delta t}{\rho R} \ ,
	\end{equation}
	where $a_{stag}$ is the shell acceleration at stagnation, $\Delta t$ is the confinement time and $P_{stag}$ is the hotspot stagnation pressure. We note that the Lawson criterion appears in the numerator which simplifies in the limit of negligible initial hotspot pressure (equation 17 of Hurricane \textit{et al.}\cite{Hurricane2020}):
	\begin{align}
	a_{stag} \Delta t &\approx \frac{1}{\sqrt{3}}\frac{\rho R_{ave}}{\rho R} \left(1-f^2\right) v_{imp} \ , \\
	f &\equiv \frac{v_{HS}}{v_{imp}} \ ,
	\end{align}
	where the areal densities of the thick and thin sides of the piston are given by:
	\begin{align}
	\rho R_{max} &= \rho R_{ave} (1+f) \ , \\
	\rho R_{min} &= \rho R_{ave} (1-f) \ ,
	\end{align}
	and $v_{HS}$ is the hotspot velocity and $v_{imp}$ is the implosion velocity. We can use this formula for the change in shell fluid velocity as a prediction of the fluid velocity variance contributing the apparent dense fuel temperature. As in Crilly \textit{et al.}\cite{Crilly2020}:
	\begin{align}
	T_{nT} &= \langle T_i \rangle + m_T \mbox{Var}(\vec{v}_f \cdot \hat{v}_n) \ , \\
	&\approx \langle T_i \rangle + m_T (a_{stag} \Delta t)^2 \ .
	\end{align}
	If we assume the anisotropy in the measured $T_{nT}$ is dominated by the differential acceleration then the following expression can be derived:
	\begin{align}
	\Delta T_{nT} &= \frac{1}{2}m_T\left(a_{stag,min}^2-a_{stag,max}^2\right)\Delta t^2 \ , \\
	&= -\frac{2}{3} f m_T v_{imp}^2 = -\frac{2}{3} m_T v_{imp} v_{HS} \ .
	\end{align}
	We have defined $\Delta T_{nT}$ here as half the full anisotropy as then the apparent dense fuel velocity is given by $T_{nT} = T_{nT}^{1D} \pm \Delta T_{nT}$ when measured parallel and anti-parallel to the hotspot velocity direction.

	\section*{References}
	\bibliography{MuCFRefs}
	
\end{document}